%% file: 0_main.tex
\documentclass[runningheads]{llncs}
\usepackage[T1]{fontenc}
\usepackage{graphicx}
\usepackage[disable]{todonotes} 
\usepackage{chronology}
\usepackage{booktabs} 
\usepackage{xspace}
\usepackage{subcaption}
\usepackage{multirow}
\usepackage{graphicx}
\usepackage{tikz}
\usepackage[edges]{forest}
\usepackage{hyperref}
\usepackage{array}
\usepackage{amsmath}
\usepackage{colortbl}
\usepackage{xcolor}
\definecolor{mygrey}{RGB}{240,240,240}
\definecolor{mygreen}{RGB}{0,128,0} 
\definecolor{myorange}{RGB}{255,140,0} 

\begin{document}
\title{On the Robustness of Cover Version Identification Models: A Study Using Cover Versions from YouTube}
\titlerunning{On the Robustness of Cover Version Identification Models}
%
\author{Simon Hachmeier\orcidID{0000-0003-4843-5196} \and
Robert Jäschke\orcidID{0000-0003-3271-9653}}
\authorrunning{Simon Hachmeier and Robert Jäschke}
%
\institute{Berlin School of Library and Information Science, Humboldt-Universität zu Berlin, Berlin, Germany
\\ \email{simon.hachmeier@hu-berlin.de}, \email{robert.jaeschke@hu-berlin.de}}
\maketitle              
%

\input{0_macros}

\begin{abstract}
Recent advances in cover song identification have shown great success. However, models are usually tested on a fixed set of datasets which are relying on the online cover song database SecondHandSongs.
It is unclear how well models perform on cover songs on online video platforms, which might exhibit alterations that are not expected. In this paper, we annotate a subset of songs from YouTube sampled by a multi-modal uncertainty sampling approach and evaluate state-of-the-art models. We find that existing models achieve significantly lower ranking performance on our dataset compared to a community dataset. We additionally measure the performance of different types of versions (e.g., instrumental versions) and find several types that are particularly hard to rank. Lastly, we provide a taxonomy of alterations in cover versions on the web.\rv{Angepasst wegen RV1: (1) The abstract needs to include more specific and interesting results.}

\keywords{cover \and song \and music \and retrieval \and longtail}
\end{abstract}

%
%
\input{1_intro}

\input{2_related}

\input{3_creation}

\input{5_analysis}
\input{6_benchmark}

\input{7_conclusion}

\begin{credits}
\subsubsection{\ackname} 
We thank the music experts which helped us in our dataset curation process. 
\end{credits}
%
%
%
\bibliographystyle{splncs04}
\bibliography{literature}






\end{document}

%% file: 0_macros.tex

\definecolor{rv}{rgb}{0.998,0.722,0.635} 
\definecolor{sha}{rgb}{0.635,0.998,0.722} 

\definecolor{rja}{rgb}{0.878, 0.831, 0.482} 
\definecolor{msc}{rgb}{0.721, 0.576, 0.862} 
\definecolor{far}{rgb}{0.576, 0.862, 0.592} 
\definecolor{hsa}{rgb}{0.576, 0.788, 0.862} 
\definecolor{mpa}{rgb}{0.978,0.534,0.534} 

\definecolor{TODO}{rgb}{0.784,0.145,0.00}

\newcommand{\pkomm}[3][TODO]{\todo[color=#1,size=\scriptsize,#2]{\sffamily #1: #3}}

\newcommand{\rv}[1]{\pkomm[rv]{inline}{#1}}\newcommand{\mffi}[1]{\pkomm[rv]{noinline}{#1}} 
\newcommand{\sha}[1]{\pkomm[sha]{inline}{#1}}\newcommand{\msha}[1]{\pkomm[sha]{noinline}{#1}} 
\newcommand{\rja}[1]{\pkomm[rja]{inline}{#1}}\newcommand{\mrja}[1]{\pkomm[rja]{noinline}{#1}} 
\newcommand{\msc}[1]{\pkomm[msc]{inline}{#1}}\newcommand{\mmsc}[1]{\pkomm[msc]{noinline}{#1}} 
\newcommand{\far}[1]{\pkomm[far]{inline}{#1}}\newcommand{\mfar}[1]{\pkomm[far]{noinline}{#1}} 
\newcommand{\hsa}[1]{\pkomm[hsa]{inline}{#1}}\newcommand{\mhsa}[1]{\pkomm[hsa]{noinline}{#1}} 

\newcommand{\final}[1]{\textbf{/* for camera ready/long version: #1  */}}

\newcommand{\eg}{e.g.,\xspace}
\newcommand{\ie}{i.e.,\xspace}
\newcommand{\etc}{etc.\xspace}
\newcommand{\csi}{VI\xspace}
\newcommand{\csilong}{Version Identification\xspace}
\newcommand{\shslong}{SecondHandSongs\xspace}
\newcommand{\shsshort}{SHS\xspace}
\newcommand{\disagraud}{DisagrAudio\xspace}
\newcommand{\disagrtxt}{DisagrText\xspace}
\newcommand{\mutunc}{MutualUnc\xspace}
\newcommand{\datacos}{Da-Tacos\xspace}
\newcommand{\oodlong}{out-of-distribution\xspace}
\newcommand{\oodshort}{OOD\xspace}
\newcommand{\shsK}{SHS100K\xspace}
\newcommand{\shsyt}{SHS-YT\xspace}
\newcommand{\shsytB}{SHS-YT+2Q\xspace}
\newcommand{\shsseed}{SHS-SEED\xspace}
\newcommand{\shsseedyt}{SHS-YT+AllQ\xspace}
\newcommand{\ytcrawl}{YT-CRAWL\xspace}
\newcommand{\yt}{YouTube\xspace}

\newcommand{\ytmatch}{YT-Match\xspace}
\newcommand{\ytversion}{YT-Positive\xspace}
\newcommand{\shsversion}{SHS-Positive\xspace}
\newcommand{\ytnonversion}{YT-Negative\xspace}
\newcommand{\shsnonversion}{SHS-Negative\xspace}
\newcommand{\ytmatchs}{YT-Matches\xspace}
\newcommand{\ytversions}{YT-Positives\xspace}
\newcommand{\shsversions}{SHS-Positives\xspace}
\newcommand{\ytnonversions}{YT-Negatives\xspace}
\newcommand{\shsnonversions}{SHS-Negatives\xspace}

\newcommand{\ytnomusic}{YT-NoMusic\xspace}
\newcommand{\match}{Match\xspace}
\newcommand{\version}{Version\xspace}
\newcommand{\nonversion}{NonVersion\xspace}
\newcommand{\nomusic}{NoMusic\xspace}
\newcommand{\ambclass}{uncertainty class\xspace}
\newcommand{\ambclasses}{uncertainty classes\xspace}

\newcommand{\ytb}[1]{\href{https://youtu.be/#1}{#1}}

\setcounter{topnumber}{5}
\setcounter{bottomnumber}{5}
\setcounter{totalnumber}{20}
\renewcommand{\topfraction}{1}
\renewcommand{\bottomfraction}{1}
\renewcommand{\textfraction}{0}
\renewcommand{\floatpagefraction}{0.99}



%% file: 1_intro.tex
\section{Introduction}

In the context of western popular music a cover version is a derivative of an original performance of a musical work. Artists perform versions to convey their subjective interpretations of musical works, which is a long-standing practice in musical culture. Usually, different versions of the same work share similar changes of individual notes (melody) or groups of notes (harmony) over time \cite{yesiler2021audio}. 

The research field of version identification (\csi) deals with the automatic detection of cover versions in music collections. Recent approaches in \csi aim to encode versions into representations retaining only relevant information in the context of cover versions \cite{bytecover2,bytecover3,bytecover,lyracnet,liu2023coverhunter,move,re-move}. For instance, Abrassart and Doras \cite{what_if} report that melody, harmony and lyrics are generally more relevant than rhythm. However, the actual relevance of each characteristic is non-trivial to predict and might strongly vary for different musical pieces. In contrast, characteristics irrelevant in the \csi context are usually well agreed upon; such as the tempo or the key/scale \cite{bytecover3,bytecover,liu2023coverhunter,yesiler2021audio,move,yu2020learning}.

Online video platforms feature various application scenarios for \csi such as copyright infringement detection and music recommendation. 
Hence, the robustness of methods against noise and variance on the platform is important. One key peculiarity of \csi in online videos is the alignment problem. In \csi, this was addressed by summarization of musical content along the time axis including pooling mechanisms \cite{bytecover2,bytecover,move,yu2020learning}
and more recently by the matching of smaller chunks of the pairs \cite{bytecover3,liu2023coverhunter}. Since \yt is a collection of videos rather than versions,\footnote{Except for YouTube's proprietary music streaming service \emph{YouTube Music}.} the relationship between videos and versions is an $m$-to-$n$ relationship. This makes the alignment problem in online videos particularly challenging. For instance, a video might contain multiple versions (\eg concert recordings) or only chunks of a version (\eg guitar solo covers or tutorials \cite{hanson2018assessing}). Additionally, videos might include noise such as commentary (\eg people reacting to music \cite{mcdaniel2021popular}). Beside the alignment problem, other challenges might arise for \csi in online videos such as the absence of the main melody (\eg karaoke or instrumental versions \cite{youtube_copyright_infringement,classify_derivative}), low fidelity in amateur recordings and versions occurring only in the background as accompaniment 
\cite{martet2016circulation}. 

\csi research has made great progress in recent years, mainly measured in metrics from MIREX\footnote{See \url{https://www.music-ir.org/mirex/wiki/2021:Audio_Cover_Song_Identification}} and reported on community datasets like \shsK-Test \cite{xu2018keyinvariant} and Da-Tacos \cite{yesiler2019datacos}. However, both of these datasets are based on the platform \shslong(\shsshort)\footnote{\url{https://secondhandsongs.com/}} curated by a community of volunteers\footnote{\url{https://secondhandsongs.com/page/About}} which makes present cover song collections subject to the selection policies of the platform. For example, \emph{web covers} are considered an individual category of versions characterized by being released non-commercially.\footnote{See \url{https://secondhandsongs.com/page/Guidelines/Entities/WebCover}} At the same time, they appear to be less relevant for collaborators, since the amount of web covers is usually much lower than for commercially released covers as can be seen for the example ``Enter Sandman'' by Metallica.\footnote{See \url{https://secondhandsongs.com/work/6616}} What is more, due to a technical limitation of the application interface of \shsshort, the created datasets do not actually contain \emph{web covers}. This poses the question whether \csi models trained and evaluated on data from \shsshort are considering all relevant characteristics of versions and motivates our first research question:

\begin{description}
    \item[RQ1] Do cover version datasets based on the platform \shslong represent the distributions of cover versions and their characteristics on \yt?
\end{description}

We assume that there exists a subset of versions with specific characteristics on \yt which are relevant in the context of \csi but not found on the platform \shsshort: \oodlong data. Consequently, recent \csi models are neither trained nor evaluated on data with regard to these characteristics. We therefore propose our second research question:

\begin{description}
    \item[RQ2] Which characteristics of versions drive the uncertainty of existing \csi models?  
\end{description}

In this paper, we aim to explore the success and the challenges of \csi on \oodlong data. Rather than relying on the cover version collection \shslong, we leverage the richness of creativity of the YouTube community. Applying a multi-modal uncertainty sampling approach, we identify the most uncertain version candidates. Subsequently, we obtain human annotations by workers on the crowdsourcing platform Mechanical Turk (MTurk). Lastly, two music experts curate a subset of the dataset and provide annotations of uncertainties in the problem context, together with a taxonomy of these. 

In summary, the main contributions are:

\begin{itemize}
    \item we provide a benchmark dataset \shsyt\footnote{\url{https://github.com/progsi/SHS-YT}} created with a multi-modal uncertainty sampling approach followed by human annotations. It includes labels on an ordinal scale to reflect the complexity of \csi on online video platforms (\eg videos without musical content and identical audio tracks).
    \item two experts curate the provided dataset to gather insights into uncertainties in the \csi context of online video platforms. We also provide a taxonomy extending an existing one \cite{yesiler2021audio}.
    \item our benchmarks show that even the current state-of-the-art model under-performs on our proposed dataset. Additionally, we identify challenging alterations such as the isolation of single instruments or the vocal track which would be better addressed in the field of query-by-humming. This uncovers potential boundaries of cover version definitions.
\end{itemize}

\rv{Überleitung rausgenommen, wegen RV1: (3) The last paragraph in the introduction is unnecessary for a conference paper.}

\rv{RV2: 2. The model is primarily trained on Western popular music, and thus it could not be as directly implemented other genres. this limitation is mention in manuscript. However, I’m curious if the model has the capability to detect tune similarity in songs across different languages or musical traditions./Could  model be effective in identifying borrowed melodies or copied tunes in non-Western music or languages not included in its primary training set?"}


%% file: 2_related.tex
\begin{table*}
  \centering
  \caption{Popular \csi benchmark datasets and the seed dataset and our annotated datasets in bold text.}\label{tab:csi_benchmark_sets}
  \setlength{\tabcolsep}{5pt} 
  \begin{tabular}{@{}lrrl@{}}
    \toprule
    \textbf{Dataset} & \textbf{Works} & \textbf{Versions}  & \textbf{Remarks} \\
    \midrule
    Covers80 \cite{covers80} & 80 & 160 & Provides raw audio files. \\
    \datacos \cite{yesiler2019datacos} & 1,000 & 15,000  &  Provides extracted audio features.  \\
    Discogs-VI-YT-Test \cite{araz2024discogsvi} & 9,878 & 116,197 & Based on metadata of Discogs. \\ 
    \shsK-Test \cite{xu2018keyinvariant} & 1,692 & 10,547  & \\ 
    YouTubeCovers \cite{youtube_covers} & 50 & 350 & Currently unavailable. \\
    \textbf{\shsseed}  & 100 & 2,404 & A subset of versions from \cite{xu2018keyinvariant}. \\
    \textbf{\shsyt}  & 100 & 900 & Includes 513 rich annotations.  \\
    \textbf{\shsytB}  & 100 & 1,092 & \shsyt with \emph{two} versions from \cite{xu2018keyinvariant} per work.  \\
    \textbf{\shsseedyt}  & 100 & 3,289 & \shsyt with \emph{all} versions from \cite{xu2018keyinvariant} per work.  \\
    \bottomrule
  \end{tabular}
\end{table*}

\begin{section}{Related Work}\label{sec:related}

\subsection{\csilong}\label{sec:csi_datasets}

\csi datasets are composed of versions which are grouped by musical works. During training, \csi models are optimized to encode audio representations of versions of the same work as similar and versions of different works as dissimilar. In the evaluation scenario, each version represents a query at a time and the remaining versions are ranked based on the musical similarity computed by the \csi model. The resulting $N$ rankings for a dataset with $N$ versions serve as the input to compute retrieval metrics such as the mean average precision (MAP). 

We provide an overview of the most popular datasets in \csi which are used for benchmarking compared to the datasets used in this paper in Table~\ref{tab:csi_benchmark_sets}. Recent \csi approaches 
\cite{bytecover2,bytecover3,bytecover,lyracnet,liu2023coverhunter,move,re-move} achieve MAP scores up to 0.96 on YouTubeCovers \cite{youtube_covers} and Covers80 \cite{covers80}. The results on the larger datasets \shsK \cite{xu2018keyinvariant} and the \datacos benchmark subset \cite{yesiler2019datacos} are lower; for instance CoverHunter \cite{liu2023coverhunter} achieves the highest MAP but still does not surpass 0.90. 

A commonality of all of these datasets but Covers80, is their utilization of \shsshort as a data source. The same accounts for the respective training sets of \csi models: \shsK-Train and the training subset of \datacos \cite{move} which were used to train the recent \csi models. Consequently, versions in the dataset can be found on \yt, but are only included if these are manually collected by the \shsshort community. The question remains whether the distribution of variance of versions on \yt is appropriately represented in existing benchmark datasets. A newer dataset, namely Discogs-VI-YT, is based on Discogs\footnote{See \url{https://www.discogs.com/}} rather than \shsshort. It is the currently biggest dataset in \csi. Since it is rather new, there are currently no benchmarks of the state-of-the-art \csi models.

\subsection{Music on YouTube}

\begin{table}
  \centering
  \caption{Classes and examples of versions on \yt.}\label{tab:video_classes}
  \setlength{\tabcolsep}{5pt} 
  \begin{tabular}{lll}
    \toprule
    \textbf{Class} & \textbf{Example}  & \textbf{Discussed in} \\
    \midrule
    Official  & Official Music Video &  \cite{airoldi2016follow,youtube_user_engagement,martet2016circulation,classify_derivative,yesiler2021audio}  \\
         & Professional Live Video &   \\
    \rowcolor{mygrey}
     User-Appropriated & Lyric Video & \cite{airoldi2016follow,youtube_user_engagement,martet2016circulation} \\ 
    \rowcolor{mygrey}
     & Slideshow  &  \\ 
    Cover & Guitar Cover &  \cite{airoldi2016follow,youtube_user_engagement,classify_derivative,yesiler2021audio}  \\ 
     & Parody &  \\ 
      & Karaoke Version &  \\ 
    \rowcolor{mygrey}
    Other & Tutorial & \cite{hanson2018assessing} \\ 
    \rowcolor{mygrey}
     & Reaction Video & \cite{mcdaniel2021popular} \\ 
    \bottomrule
  \end{tabular}
\end{table}

Various studies address the richness and diversity of versions on \yt and corresponding classes. In Table~\ref{tab:video_classes} we distinguish between 4 classes of versions and provide some examples found in existing research.

Liikkanen and Salovaara \cite{youtube_user_engagement} state that music is the most popular content type on \yt. The results were derived from data about \yt search trends, the most popular videos, and channels. The authors established twelve subclasses of versions segmented into three main classes: official (uploaded by copyright owners), user-appropriated (uploaded by fans) and derivative (\eg cover versions). While the first two classes are expected to contain highly similar audio, the third class rather relies on music fans and hobby musicians. It includes stronger changes in musical characteristics; for instance covers on instruments, parodies or remixes. 

The category of user-appropriated versions is also discussed by Mertet \cite{martet2016circulation}. The author also includes a new perspective on versions, including videos which contain versions rather as an accompaniment (\eg for movie trailers). 

From an application-driven perspective, studies have implemented pipelines to cope with copyright infringement detection \cite{youtube_copyright_infringement} and music retrieval \cite{cross_modal_music_retrieval,classify_derivative} on \yt. Smith et al. \cite{classify_derivative} propose an approach processing audio, text and video features to predict a version class. Similar to \cite{youtube_follow_algorithm,youtube_user_engagement}, the authors established classes like remixes and tutorials beside official music videos and live performances. Another approach to model classes of versions on \yt derived clusters of categories of versions by a network analysis \cite{youtube_follow_algorithm}. The results emerged clusters corresponding to musical genres and situational contexts (\eg covers and tutorials).

While the classes of versions in all of these studies might be relevant for \csi research, their consideration in the field is rather limited. Yesiler et al. \cite{yesiler2021audio} construct a taxonomy where they also mention some classes and the corresponding alterations of musical characteristics. To best of our knowledge, no existing benchmarks of \csi models investigated the impact of the mentioned alterations on model robustness.

\end{section}

%% file: 3_creation.tex
\section{Dataset Creation}
\label{sec:creation}

\begin{figure*}
\centering
\includegraphics[scale=0.22]{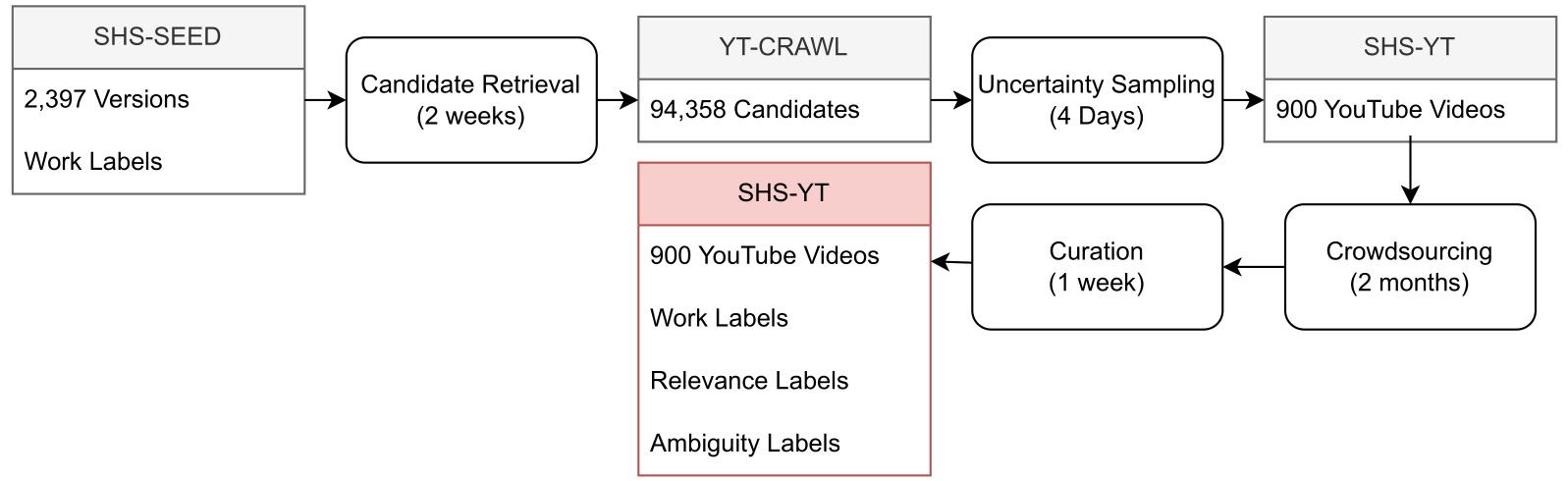}
\caption{Dataset creation. \rv{Ist das ausreichend? Wegen RV2: 1. It would be helpful to include  the timeline for database building  ( time taken by each step) , and also how much time the model takes to identify different versions of a query video.
}} \label{fig:dataset_creation}
\end{figure*}

We here describe the steps of the creation process of the dataset \shsyt as shown in Figure~\ref{fig:dataset_creation}.
We aim to evaluate the performance of state-of-the-art \csi models on \oodlong data. We select \yt as a rich source for a diverse set of versions, since there are no constraints for uploaders as opposed to the policies on \shsshort. 

To cover a representative subset of western popular music, we select the widely used \shsK-Test as a seed dataset composed of works of western popular music. In particular, we choose the first 100 works from its test subset.\footnote{See listing in \url{https://github.com/NovaFrost/SHS100K/blob/master/SHS100K-TEST}} These works are represented by 2,859 versions of which we successfully retrieved  2,397. We denote this dataset with 2,015 unique performers as \emph{\shsseed}. 

\subsection{Candidate Retrieval}

The goal of the candidate retrieval step is to obtain a set of candidate versions to be included in our dataset. We apply an approach by Hachmeier et al. \cite{ipa_query} to formulate multiple text queries per work in \shsseed. We utilize the strings for performer and title of the first version of each work to formulate queries and additionally formulate new queries using \yt search suggestions.\footnote{The list of queries per work can be found in our repository} On average, we formulate 44 text queries per work resulting in 4,365 text queries. We retrieve metadata for the top 50 videos per query\footnote{Using \href{https://pypi.org/project/}{youtube-search-python}} and drop videos with a length of 10 minutes or more. We denote the resulting collection of 94,358 videos as \emph{\ytcrawl}. We download the audio files for all videos with a sampling rate of 22,050~Hz\footnote{Using \href{https://github.com/yt-dlp/yt-dlp}{yt-dlp}.} and extract CREMA features which we use in the next step.\footnote{Using \url{https://github.com/bmcfee/crema}.} 

\subsection{Uncertainty Sampling}
\label{sec:unc_sampling}

In the uncertainty sampling step, we aim to reduce the number of versions to a smaller subset because of two reasons: Firstly, we are limited in annotation capacity. Secondly, we aim to focus on \oodlong data and want to focus on versions with characteristics not common to be found on \shsshort. 

We leverage the modalities of audio (CREMA features) and text (\yt metadata). For both domains, we use models based on deep learning as proxies. In theory, only the audio information is necessary to determine whether two versions are associated with each other. However, we use the text-based proxy to systematically find candidates where the \csi proxy over- or underestimates the musical similarity. 

\subsubsection{Modality Proxies.} 
\label{sec:model_proxies}

We use the pre-trained model Re-MOVE \cite{re-move} as a proxy in the audio/music domain which is one of the best approaches for VI at the time of dataset creation. The model processes CREMA features, which represent harmonic and melodic progressions, and encodes these into 256-dimensional embeddings. The Cosine similarity of a pair of embeddings represents their musical similarity. For the text domain, we use the entity matching model Ditto \cite{ditto}. The model is based on BERT \cite{devlin2018bert} and encodes pairs of textual entities into BERT embeddings and predicts a binary matching confidence. 
From the \shsK train dataset we create a train, validation, and test set with a ratio of 3:1:1 as proposed in \cite{ditto} with each containing positive and negative pairs of \yt videos in a 1:4 ratio. We gather the negative pairs by randomly sampling videos from another randomly selected work. We use all of the proposed data augmentation techniques and the best performing language model (RoBerta) as described in \cite{ditto}. We apply the best model checkpoint evaluated on the test set after 50~epochs for our matching task.
Since the inclusion of \yt descriptions yielded inferior results (F1 score of 0.27 against 0.95) we solely process \yt titles and channel names. 

\subsubsection{Similarity and Matching Confidence Aggregation.}
\label{sec:rel_agg}
For each candidate in \ytcrawl we compute a similarity and matching confidence with the proxy models. Since each of the works is represented by multiple versions in \shsseed (24 on average) we must aggregate the pairwise similarities and model confidences. 
For a work $i$, a set of query versions from \shsseed $Q_i$ and a candidate version $c_{ij}$ from \ytcrawl, we compute the musical similarity $S_{m}(c_{ij})$ as the arithmetic mean of the Cosine similarities of the Re-MOVE outputs of all pairs $(c_{ij}, q)$ for $q \in Q_i$. In a preliminary experiment on the validation dataset of \shsK we validated the aggregation by the arithmetic mean as opposed to aggregation by maximum.

We further compute the textual similarity $S_t(c_{ij})$ for the same pairs as the maximum matching confidence based on Ditto. We motivate this because candidates with non-matching metadata among the queries shall not have impact on the matching decision as long as at least one candidate in $Q_i$ matches. This is especially of relevance in cases with translated version titles. For instance, the version title ``Tiempo de Verano''\footnote{Spanish for ``Time of the Summer'' or ``Summertime''.} which is potentially a substring within a \yt title might match the version title ``Summertime'' with rather low confidence. Based on the aggregated values $S_{m}(c_{ij})$ and $S_t(c_{ij})$ for all candidates in \ytcrawl, we conduct uncertainty sampling with two approaches: disagreement sampling and mutual uncertainty.

\subsubsection{Disagreement Sampling.}

We establish two disagreement groups: \emph{\disagraud} denotes the candidates where the musical similarity high in contrast to the textual similarity and \emph{\disagrtxt} represents the contrary case. We measure the disagreement as the absolute difference as shown in Table~\ref{tab:uncertainty_groups} and select the three candidates with the highest disagreement for both disagreement groups per work. 

\begin{table}
    \centering
    \caption{Uncertainty groups and their constraints. We sample the top three results returned by each ranking function.}
    \label{tab:uncertainty_groups}
    \
    \begin{tabular}{@{}l@{\quad}l@{}}
    \toprule
    \textbf{Group} & \textbf{Ranking Function} \\
    \midrule
    \disagraud & $S_m(c_{ij}) - S_t(c_{ij})$\quad if\quad $S_m(c_{ij}) > S_t(c_{ij})$ \\
    \disagrtxt & $S_t(c_{ij}) - S_m(c_{ij})$\quad if\quad $S_t(c_{ij}) > S_m(c_{ij})$ \\
    \mutunc & $-\|(S(c_{ij}), S^*(C_i))\|_2$ \\
    \bottomrule
    \end{tabular}
     \end{table}

\subsubsection{Mutual Uncertainty.}
We denote the mutual uncertainty group by \emph{\mutunc} containing the top three candidates with the highest mutual uncertainty. Works with less than three candidates for \disagraud are filled with samples from this group as well. As shown in Table~\ref{tab:uncertainty_groups}, we compute the mutual uncertainty as the negative Euclidean distance between the two-dimensional vector $S(c_{ij}) = \left[S_m(c_{ij}), S_t(c_{ij})\right]^{T}$ and the vector $S^*(C_i)$, representing the center of uncertainty based on all candidates for the work $C_i$, defined as follows:
\begin{equation}
    S^*(C_i) = \begin{bmatrix} S^*_m(C_i)\\S^*_t(C_i) \end{bmatrix}
\end{equation}
with
\begin{equation}
    S^*_\theta(C_i) = \frac{1}{2} \left(S^{\min}_\theta(C_i)+S^{\max}_\theta(C_i)\right)
\end{equation}
where $\theta \in \{ m, t \}$ and $S^{\min}_\theta(C_i)$ and $S^{\max}_\theta(C_i)$ return the minimum and maximum of the Cosine similarities or matching confidences for all the candidates in $C_i$, respectively. In the following, we describe our annotation process of the resulting nine candidates per work. 

\subsection{Annotation}
\label{sec:annotation}

We impose an ordinal scale of classes and obtain annotations by workers from Amazon's Mechanical Turk (MTurk) and in-house experts.

\subsubsection{Relevance Classes.}
\label{sec:classes}

Prior \csi datasets solely consider the membership of a version to a work (binary label). Hence, each item in the dataset is expected to contain music. Further, the versions in the dataset of the same work are expected to be different regarding different aspects, such as tempo or timbre \cite{yesiler2021audio}. Both is not guaranteed when dealing with our retrieved candidates from \yt, since videos are not even guaranteed to contain music. We construct four classes on an ordinal scale with respect to the relevance to the query version:
\begin{description}
    \item[\nomusic:] Candidate version does not contain music and is not relevant.
    \item[\nonversion:] Candidate version is derived of a different work than the query versions and therefore not relevant.
    \item[\version:]
    Candidate version is derived of the same work as the query versions and therefore relevant.
    \item[\match:] Candidate version includes (parts of) the exact same audio as the original they are derived of (\emph{user-appropriated} videos). The version is relevant. 
\end{description}

We represent each work $i$ by a query version which is a random version from \shsseed. The goal of the annotation step is to gather annotations about the relevance between $i$ and the candidate in the set. We denote the resulting set of 900 annotated versions as \emph{\shsyt}.

\subsubsection{Crowdsourcing.}

\begin{figure}
 \centering
 \frame{\includegraphics[width=\linewidth]{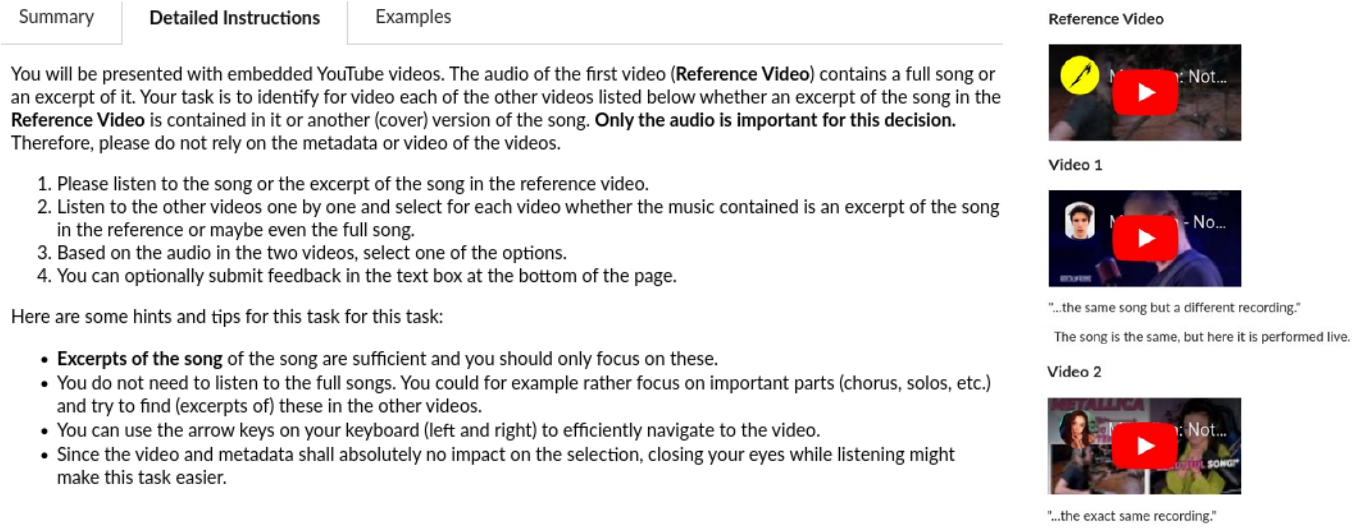}}
 \caption{Our instructions and examples to workers as presented on MTurk. Please note that the examples of the right are cropped to fit.} \label{fig:manual_worker}
 \end{figure}

We publish one human intelligence tasks (HITs) on MTurk per work with instructions and examples as shown in Figure~\ref{fig:manual_worker}. Each contains the query version, the nine candidates and a quality check candidate with a known answer (either \emph{\version} or \emph{\nonversion}) based on the works and versions in \shsseed. To simplify the task, we specifically instruct that excerpts are sufficient (\eg a medley is a \emph{\version} if it contains an excerpt of the query).
The interface and manual presented to the workers can be found in our published dataset. 

We measured the average time effort per annotation pair at 90~seconds and thus expect 15~minutes per HIT. We pay a reward of 3.2~US dollars per HIT corresponding to our domestic minimum wage, compensating our estimated time effort in consideration of the average currency exchange rate between our currency and the US dollar at annotation time.

We collect assignments by up to five workers per HIT. Following best practices to achieve annotation quality \cite{beginners_guide_mturk,mturk_positivity_bias,mturk_utility_addictionresearch,workerreputation_mturk} we only permit workers with more than 100 approved HITs and approval rate above or equal to $99\%$. We reject assignments where workers fail the quality check or complete the assignments in less than ten seconds. In some cases, we accept assignments with failed quality checks due to proper justifications by workers. We do not include these assignments in our dataset. The final worker labels are obtained by majority voting: minimum three equal labels determine the decision for the label. Candidates which remain without a final label due to high variance in label responses are curated by the experts in the next step.

\subsubsection{Curation.}

\sha{more explanation of musical terms (harmony,melody etc.)}

We employ two music experts for curation of the annotated dataset.\footnote{The two experts have 15 years of musical experience on harmonic instruments.} The experts task is to check the relevance labels of the workers for correctness, decide for a relevance label in undecided cases and to annotate the most prominent reason which makes a candidate more difficult to annotate (\ambclass). 

In cases of uncertainty, the experts discuss decisions. Ultimately, experts and authors agreed upon including boundary cases as versions as well (\eg remixes).    
The first expert curates candidates labeled with \emph{\nomusic} and 167 candidates with failed majority votes due to ties or shortage of worker assignments (because of failed assignment quality checks). 
Based on the collected reasons for uncertainties by the first expert, we formulate \emph{\ambclasses} and distinguish between uncertainties related to the version itself (\eg \emph{Song: Instrumental}) and uncertainties related to the \version in context of its occurrence in an online video (\eg \emph{Video: With Non-Music}). Some \ambclasses just apply to one relationship class, for example, \emph{Song: Same Artist} only applies if the candidate is a \emph{\nonversion}. We provide a full documentation in our published repository.\footnote{See \url{https://github.com/progsi/SHS-YT}}

The second expert utilizes the uncertainty classes directly and curates all candidates labeled with \emph{\version} and the 96 most similar candidates labeled with \emph{\nonversion}\footnote{Measured in mean Cosine similarity per benchmarking model as explained in Section~\ref{sec:rel_agg} for CQTNet and CoverHunter as explained in Section~\ref{sec:benchmark}.} for error analysis. New uncertainty classes are collected and iteratively formulated, resulting in a total of 19 ambiguity classes. Based on these classes derived of observed examples, we construct a taxonomy of alterations.

%% file: 5_analysis.tex
\section{Dataset Analysis}
\label{sec:analysis}

\subsection{Overview}

\begin{figure}
    \centering
    \subfloat[Duration in seconds. ]{{\includegraphics[width=0.48\linewidth]{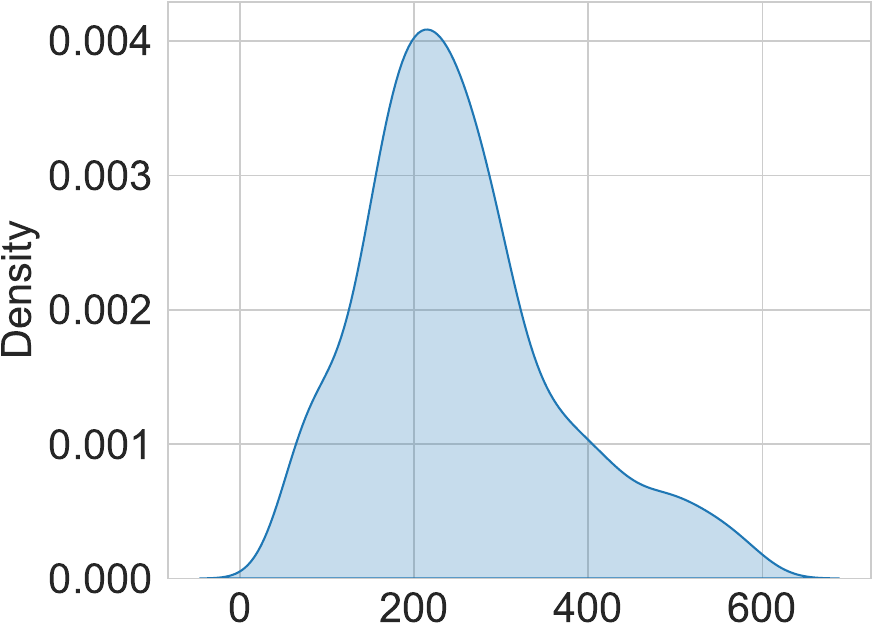}}}
    \hfill
    \subfloat[Upload date.]{{\includegraphics[width=0.48\linewidth]{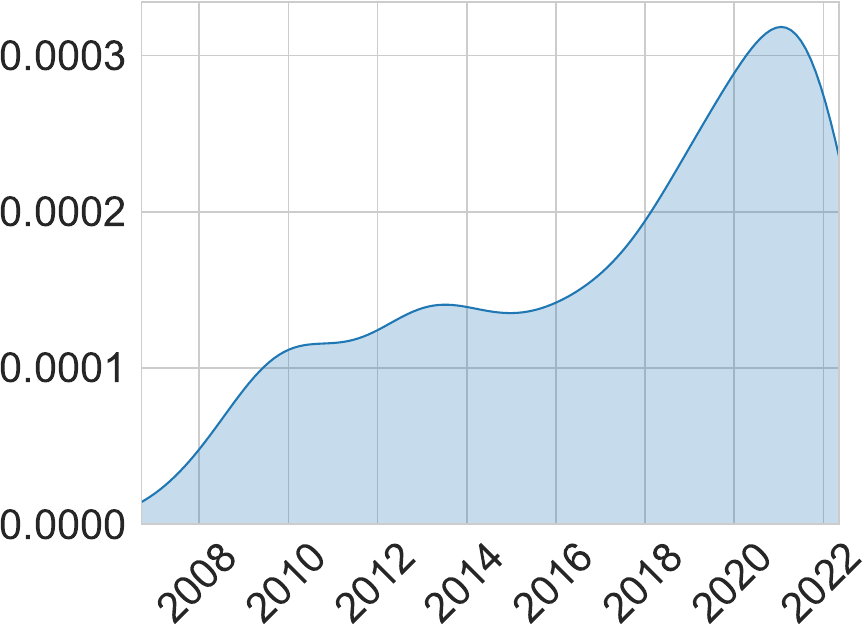}}}
    \caption{Gaussian kernel density estimates for properties of the videos in the S\shsyt dataset. The bandwith parameter is estimated by Scott's method.}
    \label{fig:youtube_statistics}
\end{figure}

We present the distributions of numerical \yt attributes of \shsyt in Figure~\ref{fig:youtube_statistics}. We observe a strong peak in duration around 3.5 minutes and in uploading dates between 2020 and 2022. 

In Table~\ref{tab:groups_labels_count} we show counts per annotation class and sampling group. The dataset mostly contains versions of other works than their respective query versions but also 197 versions of the same works. \emph{\nomusic} versions mostly occurred in the \disagrtxt group which is expected, since a modeled musical similarity by Re-MOVE is rather unlikely with the absence of actual music. Similarly, the only 4 \emph{\match} versions only occur in the \disagraud sampling group. \shsyt contains 5~versions which are also contained in Da-Tacos; all are labeled with \emph{\nonversion}. Regarding \shsK, \shsyt contains 5~versions from the test subset but from other works then in \shsseed, 2 from the validation subset and 13~candidates from the training subset. All of these candidates but one are annotated as \emph{\nonversion}. 

\begin{table}
\centering
\caption{Uncertainty sampling groups, number of occurrences per candidate label, and number of curated candidates.}
\label{tab:groups_labels_count}
\setlength{\tabcolsep}{6pt} 
\begin{tabular}{@{}lrrrrr@{}}
\toprule
& \emph{\match} & \emph{\version} & \emph{\nonversion} & \emph{\nomusic} & $\sum$ \\
\midrule
\disagraud & 4 & 89 & 200 & 0 & \textbf{293} \\
\disagrtxt & 0 & 82 & 142 & 76 & \textbf{300} \\
\mutunc & 0 & 26 & 280 & 1 & \textbf{307} \\
\midrule
$\sum$ & \textbf{4} & \textbf{197} & \textbf{622} & \textbf{77} & \textbf{900} \\
Curated & 4 & 197 & 235 & 77 & 513\\
\bottomrule
\end{tabular}
\end{table}

\begin{figure}
    \centering
    \includegraphics[width=\linewidth]{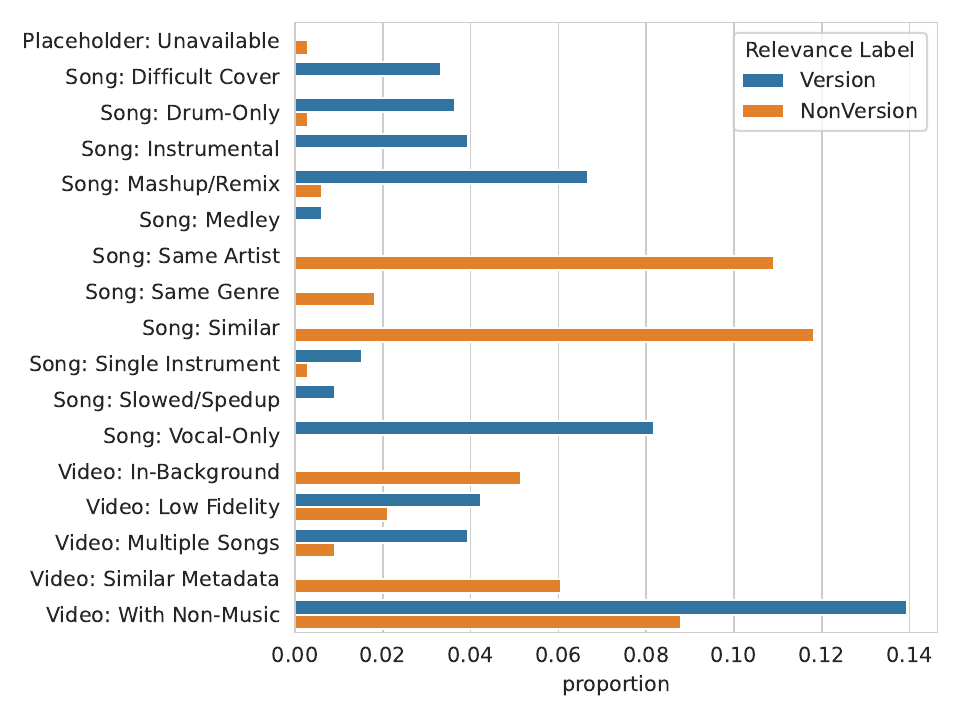}
    \caption{Relative proportion of \ambclass annotated.}
    \label{fig:cat_counts}
\end{figure}

In Figure~\ref{fig:cat_counts} we show the relative amounts of \ambclasses excluding placeholder for 104~non-ambiguous versions according to the experts. 
Non-musical content is the most represented uncertainty for \emph{\version}s with 14\% ($n=77$), followed by vocal-only. For \emph{\nonversion}s, the most frequent uncertainty is musical similarity between versions (\emph{Song: Similar}) at 12\%, followed by \emph{\nonversion}s from the same artist as the query version with 11\%.

\subsection{Annotation Quality}

Comparing the aggregated worker labels with expert labels for our 513 curated versions results in a Kendall's $\tau$ of 0.81, indicating a strong positive association. However, the agreement among workers measured in Krippendorff's $\alpha$ is just 0.43. The moderate level of inter-rater agreement might be partly due to the similarity of the \csi task to the audio music similarity task which generally is associated with limited agreement as discussed in previous studies \cite{daikoku2020human,flexer_limited_agreement_2016,flexer2019,flexer2021,jones}. Looking at the annotated \ambclasses for candidates that are falsely labeled according to the expert
($n=84$) or which did not achieve a majority vote ($n=167$) uncovers some potential issues of workers. Especially versions which include non-musical and musical content seem to confuse workers ($n=51$). We found examples from \href{https://youtu.be/aii62acsp_E}{``The Voice''} and \href{https://youtu.be/EFuBvEt84OI}{a movie scene from ``Cocktail''}. 

%% file: 6_benchmark.tex
\section{Benchmark}
\label{sec:benchmark}

\sha{RV2: Evaluation should refer to current practices about how copyright is enforced.  }

In this section, we conduct a benchmark on our proposed dataset with the goal to gather insights about the \csi performance on \oodlong data. Since \csi is a matching problem, we require relevant versions for all works in the dataset which is not the case for \shsyt. Hence, we include versions from \shsseed. We construct two benchmark datasets derived of \shsyt which we also show in Table~\ref{tab:csi_benchmark_sets}. For both datasets, we exclude the versions which are included in the training and validation datasets of \shsK:

\begin{description}
    \item[\shsytB] \shsyt with the query versions used for human annotation and one additional work from \shsseed. We select the version with the lowest version identifier which either is the original version or at least an earlier version derived of the original. This dataset includes minimum two relevant versions per work. In this dataset of 1,092 versions, our annotated versions account for around 82\%. The 312 versions labeled as irrelevant (\emph{\nonversion} and \emph{\nomusic}) account for 29\% of the dataset.
    \item[\shsseedyt] Our proposed dataset with the all versions from \shsseed resulting in 3,289 versions. Here, our annotated versions account for around 27\% of all versions. The 312 versions labeled as irrelevant account for 9\% of the dataset.
\end{description}

Beside the modality proxies (see Section~\ref{sec:model_proxies}), we evaluate two other \csi models and a fuzzy matching baseline. CQTNet \cite{yu2020learning} is a \csi consisting of mainly convolutional neural networks. It processes constant-Q transform spectograms (CQTs). The current state-of-the-art model is CoverHunter \cite{liu2023coverhunter} which also processes CQTs but includes a conformer backbone \cite{gulati2020conformer} and an attention mechanism \cite{okabe2018attentive}. The model is trained with a coarse-to-fine training scheme to address the alignment problem.
Both models are trained and validated on \shsK. We use the pre-trained models provided by the authors. In contrast to the models we benchmark, the approach by Abrassart and Doras \cite{what_if}, LyraCNet \cite{lyracnet} and the ByteCover models \cite{bytecover2,bytecover3,bytecover} are not publicly available.\footnote{We experimented with an unofficial ByteCover implementation, but it does not achieve the results which are reported in the original paper \cite{bytecover}. See \url{https://github.com/Orfium/bytecover}}

\subsection{Overall Performance}

First, we evaluate the performance of models like in traditional \csi research and consider only the binary label (relevant or not).
We report two evaluation metrics suggested by MIREX:\footnote{See \url{https://www.music-ir.org/mirex/wiki/2021:Audio_Cover_Song_Identification}} mean average precision (MAP) and mean rank of the first relevant item (MR1). Since precision for the first 10 items is not a fair metric when having works with less than 10 relevant items, we omit this metric in our evaluation. 

In Table~\ref{tab:benchmark} we report the respective results on our benchmark datasets, \shsK-Test and \datacos. Please note that we exclude Discogs-VI-YT \cite{araz2024discogsvi}. since it was published after our experiements. Furthermore, it has to be noted that both of the evaluation metrics are sensitive to the dataset size which is not negligible (see Table~\ref{tab:csi_benchmark_sets}). 
However, smaller dataset sizes usually promote a higher MAP and even though \shsytB is a smaller dataset than the others, we observe a rather strong performance drop in MAP between -34\% (CoverHunter) and -13\% (Re-MOVE).  
The performance drop is less apparent for CoverHunter at \shsseedyt and the performance even increases compared to \shsK-Test for the other \csi models. 
While this is likely due to the larger amount of versions from \shsseed, we further look into the pairwise Cosine similarities for different pairwise relationships in the following section. 

A closing remark on the overall evaluation is the potential influence of sampling bias to the performance of Re-MOVE and Ditto, since these models are used as modality proxies in dataset creation.

\begin{table*}
\centering
\caption{Benchmark results of VI models, the entity resolution model Ditto and \emph{Fuzzy} which is the token set ratio from rapidfuzz \cite{bachmann2021maxbachmann}.} 
\label{tab:benchmark}
\setlength{\tabcolsep}{6pt} 
\begin{tabular}{lrrrrrrrrr}
\toprule
& \multicolumn{2}{c}{\shsytB} & \multicolumn{2}{c}{\shsseedyt} & \multicolumn{2}{c}{\shsK-Test} & \multicolumn{2}{c}{\datacos} \\
& MAP  & MR1 & MAP  & MR1 & MAP  & MR1 & MAP  & MR1 \\
\midrule
CoverHunter \cite{liu2023coverhunter} & \textbf{0.52} & 44.5 &  \textbf{0.83} & \textbf{8.11} & \textbf{0.86} & \textbf{11.9} & \textbf{0.85} & 12.2 \\
CQTNet \cite{yu2020learning} & 0.50 & \textbf{38.8} &  0.72 & 12.43 & 0.66 & 54.9 & 0.74 & \textbf{10.7} \\
Re-MOVE \cite{re-move} & 0.40 & 86.9 & 0.56 & 18.52 & 0.53 & 38.0 & 0.52 & 38.0 \\
\midrule
Ditto \cite{ditto} &  0.39 & 73.78 & 0.78 & 18.5 & - & - & - & - \\
Fuzzy \cite{bachmann2021maxbachmann}  & 0.24 & 101.3 & 0.46 & 14.25 & - & - & - & - \\

\bottomrule
\end{tabular}
\end{table*}

\subsection{Distributions of Cosine Similarities}

To support a more well-grounded verdict about the difference of distributions of versions in \shsyt to versions on \shsshort and hence in datasets like \shsK and \datacos, we investigate the Cosine similarities of pairs of versions. A version from \shsseed can be considered a baseline version (\shsversion). Our RQ1 aims to uncover whether existing \csi models treat two \emph{\shsversion} \ of the same work as more similar then an \emph{\shsversion} compared to a version from \shsyt of the same work (\emph{\ytversion}). Similarly, the question arises whether \emph{NonVersions} from \shsyt (\emph{\ytnonversion}) are more similar than other \emph{NonVersions} from \shsseed (\emph{\shsnonversion}): the former are versions in the same \yt result sets (\eg of the same artists) and the latter are random other versions.

In Table~\ref{tab:mean_cos_sims} we show statistics about the respective Cosine similarity distributions of \emph{\shsversions} compared to other types of versions based on the relevance class. We observe that the similarities among \emph{\shsversion}s is significantly lower than their similarity to \emph{\ytversion}. Also, similarities of \emph{\shsversions} of different works are significantly lower than their similarities to \emph{\ytnonversions}; but the corresponding effect size is lower. Both of these observations are likely a reason for less consistent rankings based on the tested \csi models and hence the lower MAP scores observed in the previous section. Additionally, these insights substantiate an answer to RQ1 that in fact there exist different distributions of versions on \shsshort and \yt.

Our imposed ordinal relevance classes also allow for analysis of similarities when dealing with highly similar versions (\emph{\ytmatch}) and versions without music (\emph{\ytnomusic}). Interestingly, the similarities are neither significantly higher nor lower than the similarities to other \emph{S\shsversions}. Regarding \emph{NoMusic} versions, we can also see rather high similarities which indicates a lack of robustness of \csi models.


\begin{table*}
    \centering
\caption{Arithmetic means and standard deviations of Cosine similarities between the \shsversions and the respective other versions. The prefix \emph{YT-} indicates that the version is from \shsyt and \emph{SHS-} indicates that it is from \shsseed.}
\begin{tabular}{lrrrr}
\toprule 
 Relevance Class & CoverHunter & CQTNet & Re-MOVE & Support \\\midrule
\rowcolor{mygrey}
  \shsversion & $0.88 \pm 0.07$ & $0.61 \pm 0.13$ & $0.62 \pm 0.16$ & 96,502 \\
 \ytmatch & $0.87 \pm 0.08$ & $0.61 \pm 0.17$ & $0.66 \pm 0.19$ & 44 \\
\ytversion & *$0.80 \pm 0.10$ & *$0.48 \pm 0.19$ & *$0.45 \pm 0.24$ & 5,021 \\
\midrule
\rowcolor{mygrey}
\shsnonversion & $0.68 \pm 0.04$ & $0.33 \pm 0.08$ & $0.36 \pm 0.09$ & 5,637,128 \\
\ytnonversion & *$0.72 \pm 0.05$ & *$0.37 \pm 0.09$ & *$0.41 \pm 0.14$ & 14,305 \\
\ytnomusic & $0.68 \pm 0.07$ & $0.23 \pm 0.05$ & $0.22 \pm 0.05$ & 1,810 \\
\bottomrule
\end{tabular}
\label{tab:mean_cos_sims}
\end{table*}

To address RQ2, we investigate the differences of Cosine similarities for subsets of relevance classes grouped by their corresponding \ambclasses in Table~\ref{tab:cos_sims_categories}. 
Almost all the \emph{\ytversion}s are significantly less similar compared to \shsversions ($p < 0.01$). The most challenging classes for all the models appear to be drum-only versions, instrumental versions and medleys. While the latter is rather attributed to an alignment problem, the other two are most likely affected by the absence of the main melody and partly the harmony. Vocal-only versions which most likely only contain the main melody, appear to be hard for CQTNet and Re-MOVE and less so for CoverHunter. Difficulties for \csi models for \emph{\ytnonversion}s appear to arise due to versions being of the same artist, genre or just because they are similar by chance (\emph{Version: Similar Version} and \emph{Version: Mashup/Remix}). 

\begin{figure}
    \centering
    \includegraphics[width=\linewidth]{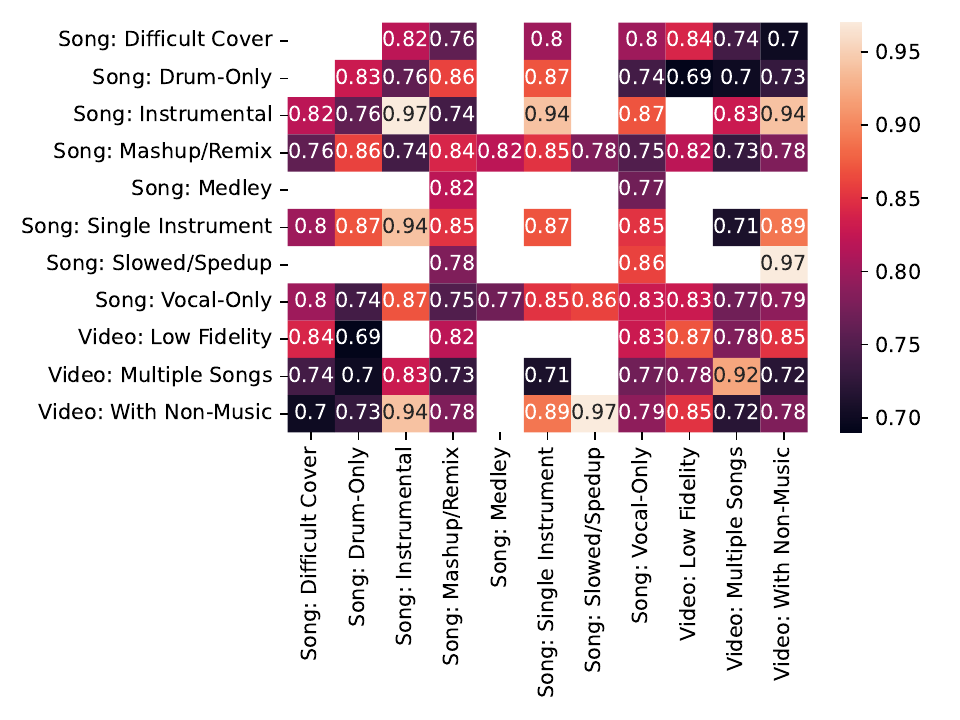}
    \caption{Mean Cosine similarities of CoverHunter embeddings between YT-Versions per respective \ambclass.}
    \label{fig:heatmap}
\end{figure}

In Figure~\ref{fig:heatmap}, we further investigate the mean similarities by CoverHunter of different relevant versions. Apparently, the difficulty of drum-only versions is validated. We can also see that versions referring to multiple versions or including non-music noise impact the similarity. In the next section, we provide some examples for versions on \yt which appear to be very challenging.

\begin{table*}
\centering
\caption{Arithmetic mean and standard deviation of Cosine similarities between versions in \shsseed and a version from \shsyt grouped by the \ambclass. *Means are statistically significant to the ones of the baseline (\shsversion or \shsversion respectively) measured with the Two-Sample-$t$-Test with $p<0.01$.}
\begin{tabular}{>{\centering\arraybackslash}p{0.1\linewidth}lrrrrrrr}
\toprule
 & Uncertainty Class & CoverHunter & CQTNet & Re-MOVE & Support \\
\midrule
\rowcolor{mygrey}
\multirow{11}{*}{\rotatebox{90}{\ytversion}} & \shsversion & $0.88 \pm 0.07$ & $0.61 \pm 0.13$ & $0.62 \pm 0.16$ & 96,502 \\
& Version: Difficult Cover & *$0.82 \pm 0.11$ & *$0.55 \pm 0.17$ & *$0.55 \pm 0.20$ & 293 \\
& Version: Drum-Only & *$0.72 \pm 0.05$ & *$0.28 \pm 0.07$ & *$0.23 \pm 0.06$ & 321 \\
& Version: Instrumental & *$0.68 \pm 0.12$ & *$0.38 \pm 0.24$ & *$0.38 \pm 0.28$ & 364 \\
& Version: Mashup/Remix & *$0.76 \pm 0.07$ & *$0.44 \pm 0.12$ & *$0.41 \pm 0.21$ & 518 \\
& Version: Medley & *$0.72 \pm 0.03$ & *$0.32 \pm 0.09$ & *$0.25 \pm 0.06$ & 86 \\
& Version: Single Instrument & *$0.80 \pm 0.05$ & $0.68 \pm 0.13$ & *$0.46 \pm 0.10$ & 195 \\
& Version: Slowed/Spedup & $0.87 \pm 0.05$ & *$0.54 \pm 0.14$ & *$0.43 \pm 0.24$ & 63 \\
& Version: Vocal-Only & *$0.77 \pm 0.04$ & *$0.38 \pm 0.09$ & *$0.23 \pm 0.07$ & 718 \\
& Video: Low Fidelity & $0.86 \pm 0.09$ & *$0.57 \pm 0.16$ & *$0.49 \pm 0.29$ & 292 \\
& Video: Multiple Versions & *$0.79 \pm 0.09$ & *$0.49 \pm 0.17$ & *$0.52 \pm 0.21$ & 343 \\
& Video: With Non-Music & *$0.81 \pm 0.10$ & *$0.48 \pm 0.18$ & *$0.50 \pm 0.23$ & 1,027 \\
\midrule
\rowcolor{mygrey}
\multirow{7}{*}{\rotatebox{90}{\ytnonversion}} & \shsnonversion & $0.68 \pm 0.04$ & $0.33 \pm 0.08$ & $0.36 \pm 0.09$ & 5,637,128 \\
& Version: Mashup/Remix & *$0.78 \pm 0.04$ & *$0.42 \pm 0.07$ & $0.33 \pm 0.14$ & 53 \\
& Version: Same Artist & *$0.76 \pm 0.05$ & *$0.45 \pm 0.08$ & *$0.51 \pm 0.11$ & 862 \\
& Version: Same Genre & *$0.75 \pm 0.05$ & *$0.40 \pm 0.09$ & *$0.53 \pm 0.13$ & 169 \\
& Version: Similar Version & *$0.76 \pm 0.06$ & *$0.45 \pm 0.08$ & *$0.51 \pm 0.11$ & 1,069 \\
& Video: Multiple Versions & *$0.71 \pm 0.05$ & *$0.39 \pm 0.08$ & *$0.40 \pm 0.14$ & 102 \\
\bottomrule
\end{tabular}
\label{tab:cos_sims_categories}
\end{table*}

\subsection{Error Analysis}

We examine the reasons for uncertainties more profoundly. First, we look at versions labeled \emph{\nonversion} which are more similar than random other versions. We found that songs of the same genres are generally more similar, for instance bossa nova and blues. In theory \csi models are not optimized to model genres per se. However, musical characteristics such as chord progressions (\eg blues scheme) or rhythm (\eg bossa nova beat) seem to be hard to disentangle from \csi representations. Similarly, musical characteristics appear to correlate for versions of the same artists (\eg Lady Gaga, Backstreet Boys, AC/DC). However, in some cases versions appear to be similar simply by similar chord progressions (\href{https://youtu.be/ZFWC4SiZBao}{``Ultraviolence''} and \href{https://youtu.be/E5sVhFnrlTw}{``Radioactive''}). Interestingly, we also found pairs of versions with high similarity according to CoverHunter even though one version is labeled with \emph{\ytnomusic}.\footnote{\ytb{PG6iJmbnOTY} (mute) and for the work ``What's going on'' and \ytb{svQD6mGDPXc} (mute) for the track ``Stairway to heaven''} We assume that this is due to the matching of mute or low energy sections in these versions with mute parts of \emph{\shsversion}s after the alignment module.

Investigating some \emph{\ytversion}s which appear to be difficult to detect, we found that vocal-only can refer either to versions with isolated voice stem by sound source separation\footnote{\ytb{cixhJpyTWko} (``Dancing in the dark'')} but also self-recorded vocal-only versions.\footnote{\ytb{24AKYyNusvs} (``Rolling in the deep'')}

\section{Discussion and Implications}
\label{sec:discussion}

We summarize the findings gathered by our created dataset \shsyt and the respective benchmarks. Regarding RQ1 we in fact confirmed a significant difference between some of the versions on \yt and the ones included in \shsshort-based datasets. Based on our ordinal relevance labels, we derive that the difficulty especially arises due to relevant versions which are difficult to detect (false negatives) rather than irrelevant versions (false positives). However, some aspects such as similarity of songs within genres, of the same artists or with similar chord progressions seem to impact the overestimated similarity. 

Looking at our dataset with annotated uncertainty classes reveals that the drum-only videos are rather challenging as well as instrumental versions. While the former do not include melody and harmony, these cases can be denoted as boundary cases. This raises the question about how a cover version is defined which is a question to be asked in musicology and maybe even of philosophical nature. Beside these rather song-specific uncertainty classes, there are also observable difficulties arising due to the alignment problem. While this is a general problem in \csi, extreme cases such as medleys, multiple versions in a video and videos with versions and non-musical content still appear to be difficult for existing models. 

\begin{figure}
\centering
\input{images/taxonomy}
\caption{Taxonomy of Cover Versions in Online Videos} \label{fig:taxonomy}
\end{figure}
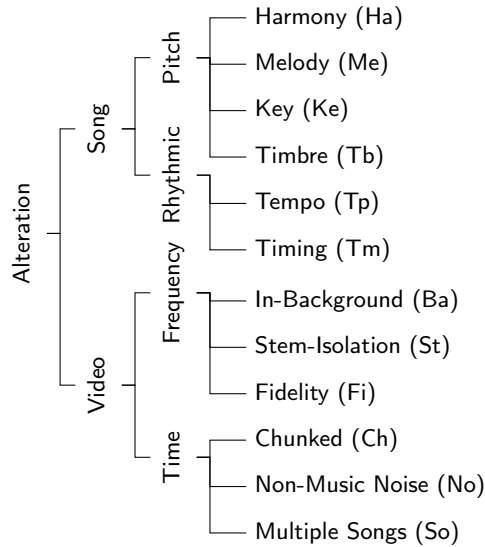

To improve \csi models in the future one solution is to rely on broader datasets in terms of data sources. For instance, by utilizing \yt metadata to train weakly-supervised models. However, we propose another solution based on our observations.  In Figure~\ref{fig:taxonomy}, we propose our taxonomy of cover versions in online videos. In the context of \csi, musical characteristics which are 
discussed by Yesiler et al. \cite{yesiler2021audio} (\emph{Song} node) are one key component to model cover version relationships. Researchers are well aware about the importance of alterations in these characteristics and address them by augmentation techniques such as pitch-variations, tempo-variations. In the context of \csi on \yt (\emph{Video} node), there are additional challenges which arise due to the context of online videos. Our observations provided examples for versions with low-fidelity and versions which occur in the background with foreground noise. We believe that both of these alterations can be well addressed by incorporating audio fingerprinting and noise mixing. We also found that isolated stems (\eg drum-only, vocal-only versions) are particularly challenging. This is a problem which points to the related music information retrieval task of query-by-humming, where audio representations rely on single stems (usually the singing voice). In \csi, an integration of sound source separation in augmentation techniques could further benefit model performance. Alternatively, rather than extracting the features in an end-to-end fashion using CQT spectrograms, one could extract features for melody, harmony and rhythm separately like Abrassart and Doras \cite{what_if}.

Lastly, the alignment problem which we have mentioned appears to be particularly strong on online video platforms. Not only can a version be represented only by a section (\emph{Chunked}), but also along with multiple other versions or non-music noise.  
The application of sliding time windows, possibly with different sizes followed by a maximum aggregation can address this problem. However, this might in turn increase the risk for false negatives and the computational load. We propose that also the synthesis of data by concatenation of different versions and non-musical noise such as commentary can help to make \csi models more robust for these cases.

%% file: images/taxonomy.tex
\usetikzlibrary{decorations.pathmorphing}

\forestset{
  label tree/.style={
    for tree={tier/.option=level},
    level label/.style={
      before typesetting nodes={
        for nodewalk={current,tempcounta/.option=level,group={root,tree breadth-first},ancestors}{if={>OR={level}{tempcounta}}{before drawing tree={label me=##1}}{}},
      }
    },
    before drawing tree={
      tikz+={\coordinate (a) at (current bounding box.north);},
    },
  },
  label me/.style={tikz+={\node [anchor=base west] at (.parent |- a) {#1};}},
  striped edge/.style={edge={decorate, decoration={snake, amplitude=1.5, segment length=4mm}}, edge path={\noexpand\path[\forestoption{edge}] (!u.parent anchor) -- ([yshift=-3pt].child anchor) \forestoption{edge label};}},
}

\begin{forest}
  forked edges,
  label tree,
  for tree={
    grow=0,  
    s sep'=0pt, 
    anchor=base west, 
    font=\strut\footnotesize\sffamily,
    edge=black, 
  },
  [Alteration, rotate=90 [Video, rotate=90 [Time, rotate=90 [Multiple Songs (So)] [Non-Music Noise (No)] [Chunked (Ch)]] [Frequency, rotate=90 [Fidelity (Fi)] [Stem-Isolation (St)] [In-Background (Ba)]]] [Song, rotate=90 [Rhythmic, rotate=90 [Timing (Tm)] [Tempo (Tp)]] [Pitch, rotate=90 [Timbre (Tb)] [Key (Ke)] [Melody (Me)] [Harmony (Ha)]]]] 
\end{forest}

%% file: 7_conclusion.tex
\section{Conclusion and Limitations}
\label{sec:conclusion}

\sha{Discuss uncertainty sampling approach}
\sha{RV2: Lack of a more general consideration about the performance of the task in terms of how social media are currently managed }

In this paper we proposed \shsyt, a new benchmark dataset for VI. Created with a multi-modal uncertainty sampling approach and annotated by workers and experts including \ambclasses, this dataset provided novel insights in the robustness of \csi models. Lastly, we want to highlight some limitations of our study.

To the best of our knowledge, this is the first study which evaluates \csi approaches with regard to different alterations among versions focusing on the most prominent uncertainty. Nevertheless, these classes might be partly subjective and cannot be fully isolated by other effects which might occur for certain pairs of versions. 

Due to the peculiarity of \yt of being a dynamic online video platform, we cannot guarantee the presence of our videos on the platform in the future. In our repository, we provide all the URLs investigated. Due to copyright issues, we cannot provide the raw audio but only the extracted CQT and CREMA features. This paper focused on cover versions in the context western popular music. We are well aware that other genres might incorporate other characteristics which make this study less applicable. In future studies, the consideration of other genres with different characteristics could improve to gather an even broader overview of musical reinterpretations.